\documentstyle[twocolumn,prl,aps,epsf]{revtex}

\begin{document}

\draft
%\twocolumn[\hsize\textwidth\columnwidth\hsize\csname @twocolumnfalse\endcsname
\wideabs{

\title{Phase-dependent Kondo Resonance in a Quantum Dot Connected to a
        Mesoscopic Ring }
\author{Kicheon Kang}
\address{ Department of Physics, Chonbuk National University,
            Chonju 561-756, Chonbuk, Korea}

\author{Luis Craco}
\address{Instituto de F\'{\i}sica Gleb Wataghin - UNICAMP
            C.P. 6165, 13083-970 Campinas - SP, Brazil}
\date{\today}

\maketitle
\begin{abstract}
Phase-sensitive transport through a quantum dot coupled to an 
Aharonov-Bohm ring is analyzed. In this geometry the spectral
density of states is directly related to the conductance. It is shown
that the Kondo resonance depends on the phase and on the total number 
of electrons (modulo 4) in the mesoscopic ring. The effect of the
discrete level spacing in the ring and of the coupling to the electrical 
leads is discussed.
\end{abstract}

}
%]
\narrowtext

The Kondo effect is one of the most intensively studied
topics in many-body physics~\cite{hewson93}, which describes several 
anomalous features observed in metallic systems with embedded magnetic 
impurities. Due to the recent remarkable progress
in the nano-fabrication of electronic devices it is now possible to 
investigate the Kondo effect by using nano-structures in a highly controlled 
way~\cite{gordon98,cronen98,schmid98,simmel99}. These investigations 
clearly demonstrate that the Kondo resonance provides a new channel 
for electric current through a quantum dot (QD) attached to two separate
electrodes, and gives rise to an enhanced conductance at low bias and 
temperature~\cite{glazman88,ng88,hershfeld,meir,yeyati93,konig96,kang98,craco99}.

Basically the Kondo resonance is a phase-coherent process characterized 
by a resonance phase shift of $\pi/2$ with unitary cross section for 
symmetric junctions at $T=0$~\cite{glazman88,ng88}. The phase-coherence 
of the Kondo-resonant transmission has been theoretically investigated 
by considering an Aharonov-Bohm interferometer with a QD embedded in one 
of the two arms~\cite{bruder96,davidovich97,izumida97,gerland00}, and 
confirmed in recent experiments~\cite{wiel00,ji00}. In this geometry, the 
dot was considered to be embedded in a host with continuous spectrum, and 
the transmission amplitude through the QD 
%$(t_{QD})$ 
is independent of the 
Aharonov-Bohm flux as long as the electron transmission is dominated
by the two direct paths of electron propagation~\cite{gerland00}.

\begin{figure}[htb]
\epsfxsize=3.5in
\epsffile{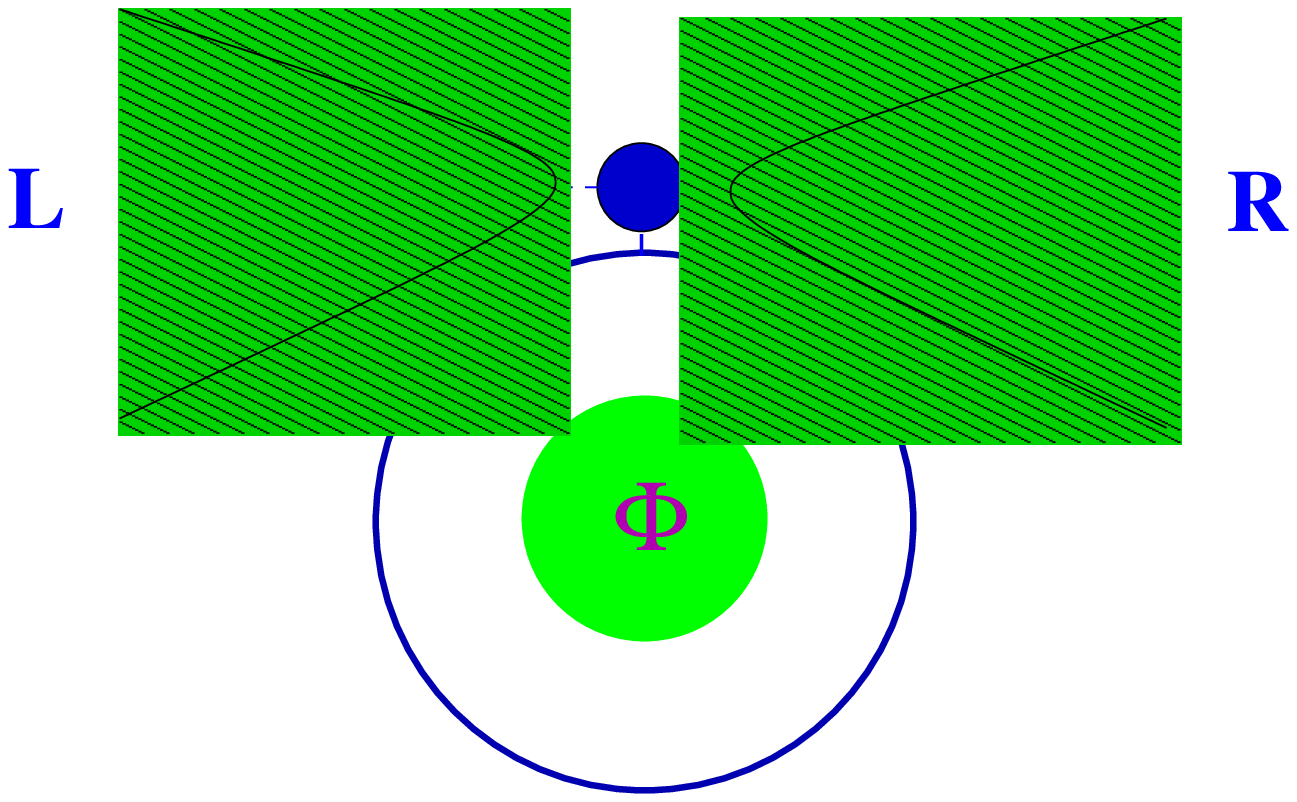}
\caption{Schematic figure of the device setup.}
\label{fig1}
\end{figure}

More recently, it has been also suggested that a {\em mesoscopic} Kondo 
effect can be detected in a QD embedded in an isolated small ring by studying 
the behavior of the persistent 
current~\cite{ferrari99,kang00,cho00,eckle00,affleck00}. 
In this case, the nature of the Kondo effect depends on the 
Aharonov-Bohm phase. Meanwhile, it seems to be a rather difficult task to 
realize experimentally such environments. In this Letter, we propose and 
theoretically investigate a device setup in which a {\em phase-dependent
Kondo resonance} can be studied by using more easily accessible transport 
measurement. The geometry considered here consists of an interacting
quantum dot, connected to one side with a Aharonov-Bohm ring and weakly 
coupled to two electrical leads (see Fig.~\ref{fig1}). 
In contrast to a system of a Kondo
impurity embedded in a mesoscopic ``box''~\cite{thimm99}, the spectral 
density of states (DOS) of the QD is directly measurable in our geometry, 
and one can systematically study mesoscopic effects on the Kondo resonance
as a function of the Aharonov-Bohm phase.

The Hamiltonian of the system can be written as

\begin{equation}
\label{eq:model}
 H = H_0 + \sum_{\alpha=L,R} H_\alpha + T \;,
\end{equation}
where $H_0$ is the Hamiltonian term for the dot-ring system, $H_\alpha$ 
describes the left (L) and the right (R) lead, and $T$ accounts for the 
tunneling between the QD and the two leads. 
$H_0$ is decomposed into
\begin{equation}
 H_0 = H_{QD} + H_{Ring} + H_{t'} \;,
\end{equation}
where $H_{QD}$, $H_{Ring}$, and $H_{t'}$ 
correspond to the Hamiltonians describing
the quantum dot, the AB ring, and the dot-ring hybridization  
at site ``0", respectively: 
\begin{mathletters}
\begin{eqnarray}
 H_{QD} &=&  \sum_\sigma \varepsilon_d d_\sigma^\dagger d_\sigma
       + U \hat{n}_\uparrow \hat{n}_\downarrow \;, \\
 H_{Ring} &=& -t \sum_{j=1}^{N}\sum_\sigma
     \left(  e^{i\varphi/N} c^\dagger_{j\sigma}
           c_{j+1\sigma} + \mbox{\rm h.c.}\right) \label{eq:hring} \;, \\
 H_{t'} &=& -t' \sum_\sigma \left( d^\dagger_\sigma
                c_{0\sigma} + c_{0\sigma}^\dagger d_\sigma \right) . 
\end{eqnarray}
\end{mathletters}
Here, we describe the ring by a tight-binding Hamiltonian with $N$ lattice 
sites, and the QD by a single Anderson impurity.  $\varepsilon_d$ 
is the single particle energy, and $U$ is the Coulomb interaction in the QD.
The phase $\varphi$ in Eq.~(\ref{eq:hring}) comes from the Aharonov-Bohm 
flux, and is defined by $\varphi = 2\pi\Phi/\Phi_0$, where $\Phi$ and 
$\Phi_0$ are the external flux and the flux quantum ($=hc/e$) respectively.  
Note that Eq.~(\ref{eq:hring}) can be diagonalized and the 
corresponding eigenvalues are 
$\varepsilon_m = -2t\cos{\frac{1}{N} (2\pi m + \varphi)}$ ($m$ being any 
integer number). The two leads are described by the corresponding 
reservoirs, consisting of non-interacting electrons with the single
particle energies $\varepsilon_{k\alpha}$ ($\alpha=L,R$):
\begin{equation}
 H_{\alpha} = \sum_{k\sigma} \varepsilon_{k\alpha}
 a_{k\sigma\alpha}^\dagger a_{k\sigma\alpha} \; ,
\end{equation}
and the difference between the chemical potentials is proportional to the
applied voltage $V$, i. e., $\mu_L-\mu_R=eV$.  Finally, the 
tunneling between the QD and the reservoirs is written as
\begin{equation}
 T = \sum_{k\sigma\alpha} \tau_k^\alpha
 \left( a_{k\sigma\alpha}^\dagger d_\sigma + \mbox{ h.c.} \right) \;.
\end{equation}

In the wide-band limit of the reservoirs the current through the QD can be 
obtained by employing the formula~\cite{meir92}
\begin{equation}
 I = \frac{2e}{\hbar} \sum_{\sigma}\int d\omega\, 
  \tilde{\Gamma}(\omega)
  \left\{ f_L(\omega)-f_R(\omega) \right\} \rho_{\sigma}(\omega) ,
            \label{eq:curr}
\end{equation}
where $\tilde{\Gamma}(\omega) = \Gamma_L(\omega) \Gamma_R(\omega) /
\Gamma(\omega)$
with $\Gamma_\alpha(\omega) = \pi\sum_k| \tau_k^\alpha |^2\,\delta(\omega
-\varepsilon_{k\alpha})$ being the coupling strength between the
QD level and the lead $\alpha$, and $\Gamma(\omega) = \Gamma_L(\omega)
+\Gamma_R(\omega)$.
The spectral density of states in the QD is
$\rho_{\sigma}(\omega)=-\frac{1}{\pi} \mbox{Im}\,G_{\sigma}(\omega)$
and $f_\alpha(\omega)=1/( e^{\beta(\omega-\mu_\alpha)}+1 )$ is the Fermi 
function of lead $\alpha$. The calculation of the current in 
Eq.~(\ref{eq:curr}) only needs the DOS in the QD, which can be obtained 
from the corresponding one-particle Green's function. 

In this work we shall calculate the retarded one-particle Green's function 
of the QD employing the following expression:
\begin{equation}
\label{eq:Gf}
G_\sigma(\omega) = \frac{1}{\omega - \varepsilon_0 
- \sum_{\alpha} \Delta_\alpha (\omega) - \eta (\omega) - \Sigma(\omega)}\;,
\end{equation}
where $\Delta_\alpha (\omega)$ 
%= \sum_{k} \tau_{k}^2/(\omega-\varepsilon_{k\alpha})
describes the self-energy contribution of the dot-leads 
tunneling~\cite{craco99,lg1},  
$\eta (\omega) = 1/N \sum_{m} t^{'2}/(\omega-\varepsilon_{m})$  
accounts for the dot-ring coupling and $\Sigma(\omega)$ is the self-energy 
contribution due to the on-site Coulomb interaction $U$. According to 
Ref.~\cite{craco99} one can address the latter problem by means of the 
following {\it ansatz}
\begin{equation}
 \Sigma(\omega) = Un + \frac{ a\Sigma^{(2)}(\omega) }{ 
  1 - b\Sigma^{(2)}(\omega) } ,
     \label{eq:ansatz}
\end{equation}
where the quantities $a$, $b$, and $n$ (the occupation number of the QD 
level) have to be determined self-consistently, and
$\Sigma^{(2)}$ is the second order self-energy in $U$.
This {\it ansatz} gives the correct weak and 
strong coupling limits for all range of parameters, including Kondo, 
charge fluctuation, and even-number site limit~\cite{yeyati93,craco99}.
We shall in what follows assume that the energy levels of the ring are 
half-filled, and consider only the symmetric case for the 
dot $(\varepsilon_d=-U/2)$. 
The former assumption implies that the number of electrons in the ring 
is equal to the number of lattice sites $(N)$.  In addition, 
we chose $U=1.2t$  and a parabolic form of 
$\Gamma_{\alpha} (\omega)$~\cite{craco99,lg1} 
centered at $\omega=0$ with bandwidth $W=4.8t$.

\begin{figure}[htb]
\epsfxsize=3.5in
\epsffile{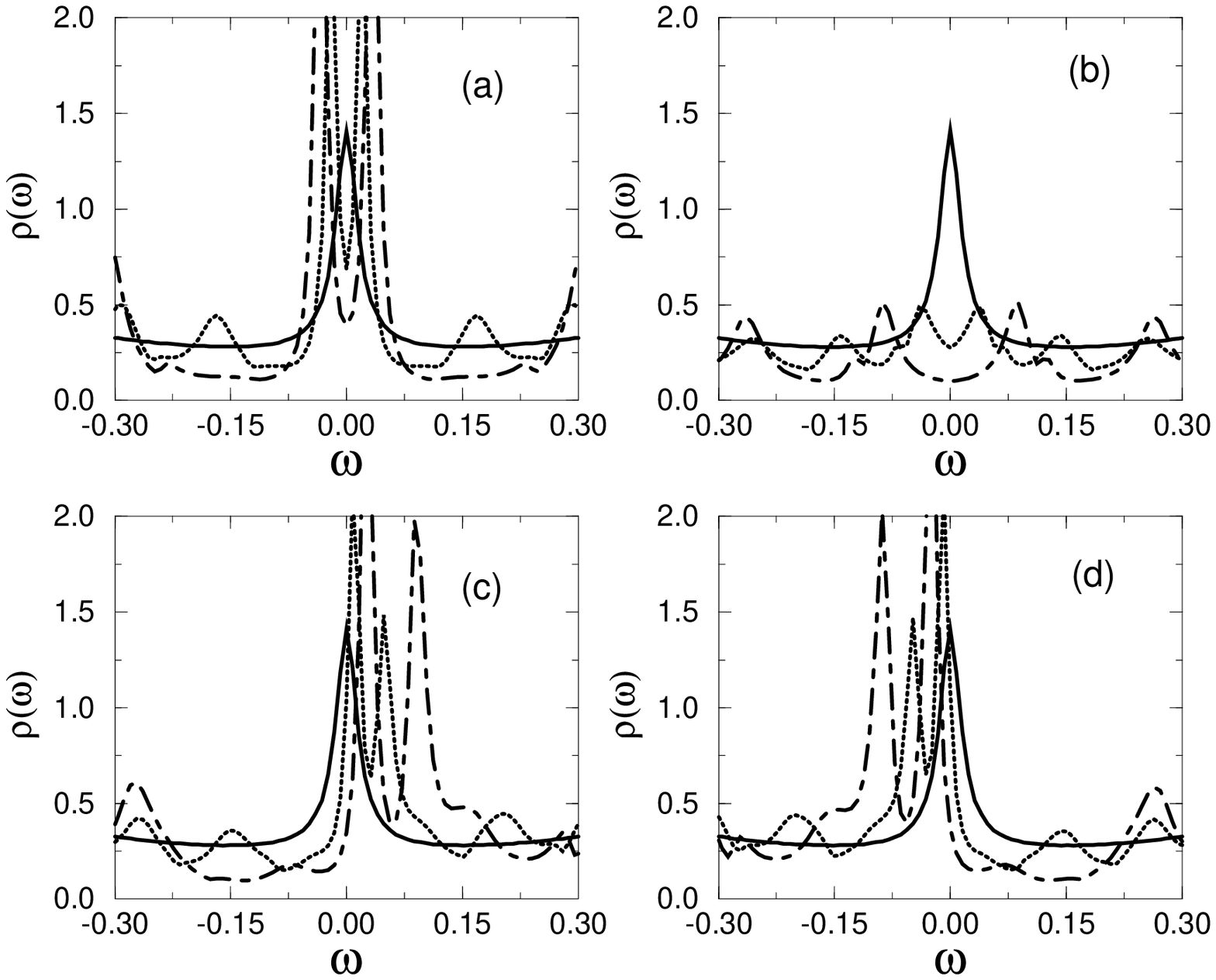}
\caption{Single-particle DOS of the QD level for $\varphi=0$ 
and the different values of $N$ (modulo 4). 
(a) N=52 (dot-dashed), N=100 (dotted) and N=1000 (solid);
(b) N=54 (dot-dashed), N=102 (dotted) and N=1002 (solid);
(c) N=53 (dot-dashed), N=101 (dotted) and N=1001 (solid); and,
(d) N=55 (dot-dashed), N=103 (dotted) and N=1003 (solid).
The remaining  parameters are $U=1.2$, $\varepsilon_d=-0.6$, 
$\Gamma'=0.08$, and $\Gamma=0$, in units of $t$.
}
\label{fig2}
\end{figure}

Let us begin by analyzing the limit of weak tunneling 
between the QD and the reservoirs $(\Gamma \ll t'^2/2t)$, where 
we shall neglect the reservoirs in the calculation of the DOS.
Moreover, to be consistent with our study of 
the effect of the reservoirs, we shall assume here a 
grand-canonical description, even for the isolated ring-dot configuration. 
Fig.~\ref{fig2} shows the $N$-dependence of the DOS for 
the QD level in the absence of the AB flux. For large  $N$, our results 
are identical for all values of $N$, and the DOS is shown to be independent 
of both $N$ and $\varphi$, as expected. It is worth mentioning that the 
DOS depends on $N_4\equiv N$ $modulo \;4$
for small sized rings ($N\lesssim 1000$ for the parameters used in 
Fig.~\ref{fig2}). Two remarkable features are found by decreasing the size 
of the ring. First, small satellite peaks appear away from the Fermi level.
These peaks are originated by the discrete level spacing in the ring, and
become pronounced by decreasing $N$. Second, the single-particle excitation 
at the Fermi level - the Kondo resonance in the continuum limit - suffers a 
drastic change. By reducing $N$, the Kondo resonance is suppressed 
($N=4n+2$) or split into two peaks with a pseudo-gap, depending on the 
value of $N_4$. Such effects are signatures of destruction of  the Kondo 
cloud, caused by  
the finite size of the ring. A similar feature has been found in the study of 
a ``Kondo box''~\cite{thimm99}, which also shows a parity-dependent 
Kondo resonance. In our case, this situation is more complicated by  
the orbital degeneracy of the ring as well as by the spin degeneracy.

\begin{figure}[htb]
\epsfxsize=3.5in
\epsffile{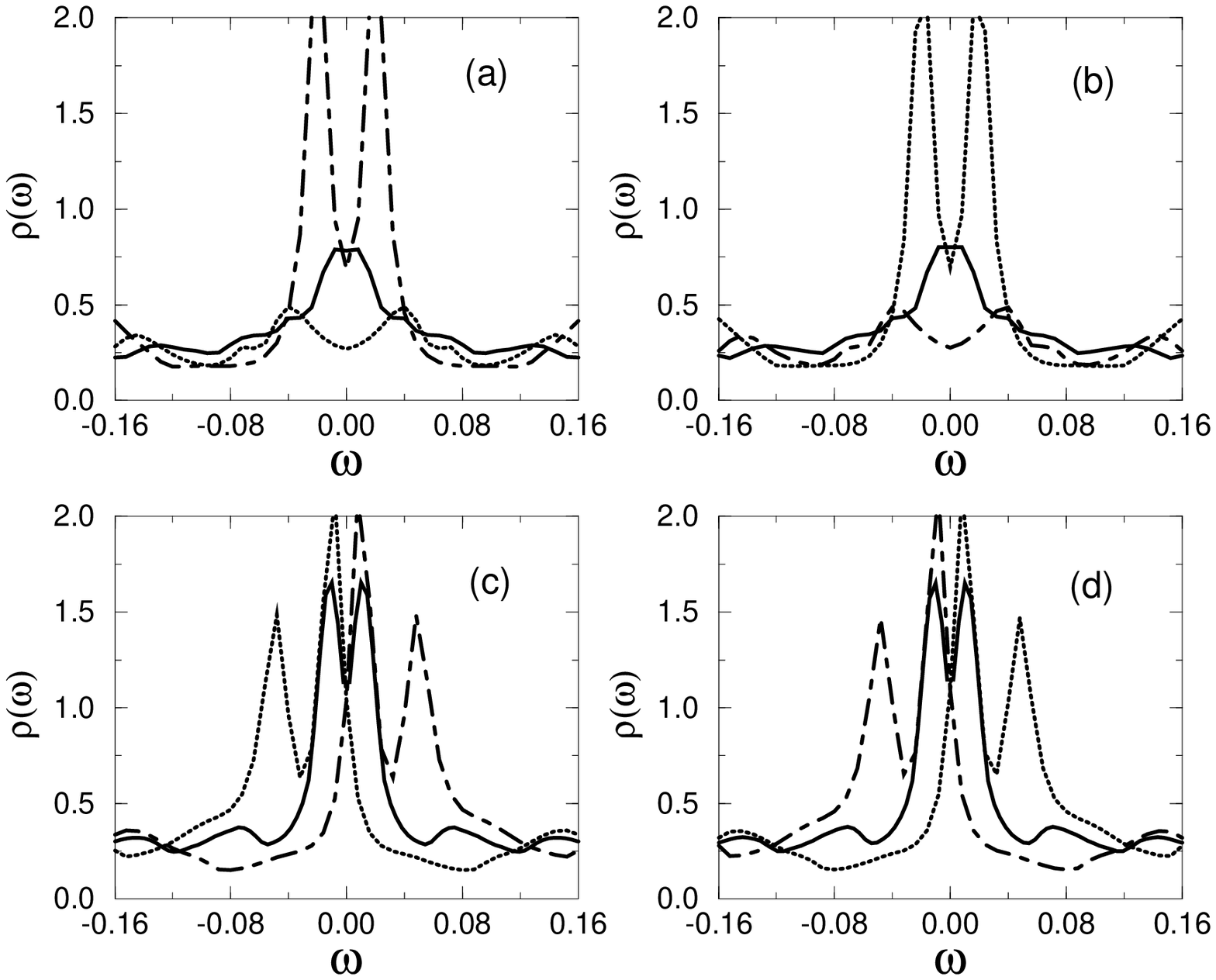}
\caption{Single-particle DOS of the QD level for
(a) N=100, (b) N=102,  (c) N=101 and (d) N=103, and different 
values of the phase: $\varphi=0$ (dot-dashed), $\varphi=\pi/2$ (solid) 
and $\varphi=\pi$ (dotted).The other parameters are the same used in Fig. 2.}
\label{fig3}
\end{figure}

Fig.~\ref{fig3} shows the phase-dependence of the Kondo resonance for 
different phases and values of $N_4$. In this figure one can see that 
the Kondo resonance 
at the Fermi level is strongly phase-dependent for small sized rings, 
and this resonance depends 
on $N_4$ as well. An interesting point to be considered is that the DOS for 
a given $N$ and $\varphi$ $(\rho_N(\omega, \varphi))$ satisfies the relation
\protect{$\rho_N(\omega,\varphi) \simeq \rho_{N+2}(\omega,\varphi+\pi)$}.
These result follows form the expression of the single
particle energies of the ring: $\varepsilon_m = -2t 
\cos{\frac{1}{N}(2\pi m + \varphi)}$, 
because for sufficiently large $N$ ($\gg1$), 
the configuration of the energy levels and its occupation near the Fermi
level is invariant under the transformation $N\rightarrow N+2$ along with
$\varphi\rightarrow\varphi+\pi$. 

\begin{figure}[htb]
\epsfxsize=3.5in
\epsffile{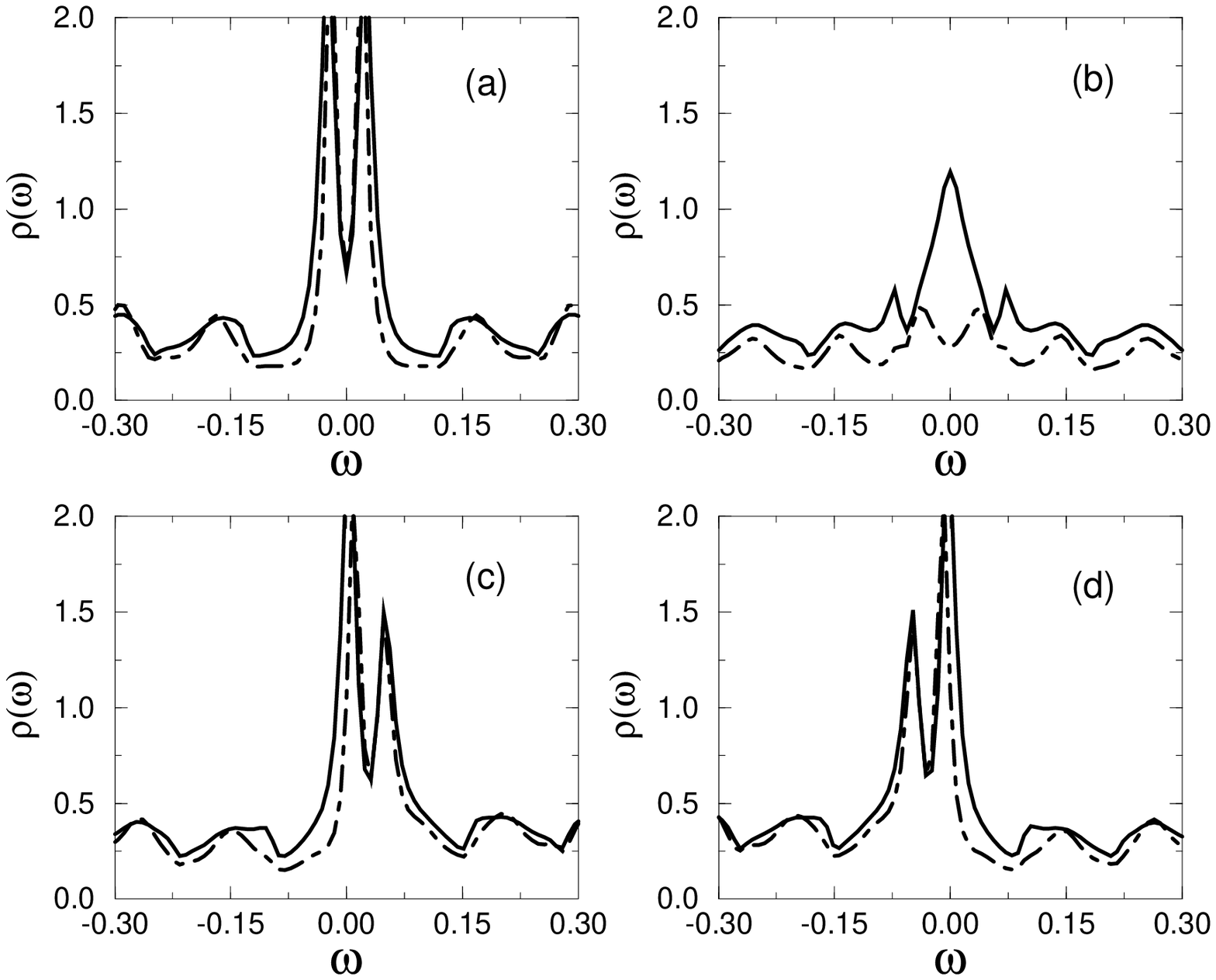}
\caption{Single-particle DOS of the QD level for
(a) N=100, (b) N=102,  (c) N=101 and (d) N=103, and different 
values for the dot-leads coupling: $\Gamma=0$ (dot-dashed) and 
$\Gamma=0.03$ (solid). The other parameters are the same used in Fig. 2.}
\label{fig4}
\end{figure}

The effects of the coupling  of the QD to the leads has been neglected 
up to here in all our calculations of the DOS, and in  Fig.~\ref{fig4} we
show those effects for small coupling ($\Gamma\ll\Gamma'$) and contiguous
values of $N$ corresponding to the four 
possible configurations. In most cases, the general feature of the DOS 
is only slightly modified due to this small coupling. As one may expect, 
the Kondo 
resonance is slightly enhanced, but for some particular 
configurations, see for example Fig.~\ref{fig4}(b), the DOS near the 
Fermi level is strongly modified, and the Kondo resonance is dominated 
by the lead-dot coupling.

In conclusion, we have studied the effect of 
an Aharonov-Bohm ring on the Kondo resonance of a mesoscopic quantum dot. 
To this purpose we have proposed and theoretically investigated a 
device setup in which the phase-dependent Kondo resonance might be analyzed 
experimentally. We have found that the Kondo resonance at the Fermi level 
of the dot is strongly affected by the size of the ring as well as  by the 
magnetic flux, which clearly demonstrates the mesoscopic nature of the 
Kondo scattering.
We have shown that the Kondo resonance is strongly suppressed or split 
into two peaks depending on the number (modulo 4) of electrons in the ring. 
It has been further  
verified that the resonance shows a strong phase-dependence for rings of small 
size. We have also analyzed the effect of the coupling between the QD and
the two leads. We show that the DOS around the Fermi level is modified by the
coupling in a manner that depends on $N_4$.

%\acknowledgements
LC wishes to acknowledge M. Foglio for useful comments.
This work was funded by the MPIPKS. 
LC was also supported by the Funda\c c\~ao de Amparo 
\`a Pesquisa do Estado de S\~ao Paulo (FAPESP). KK acknowledges support by 
Grant No. 1999-2-11400-005-5 from the KOSEF.

%%%%%%%%%%% References %%%%%%%%%%%%%%%%%%%%%%%%%%%%%%%%%%%%%%%%%%%%%%%

\end{document}